# VERSATILE CAPPILARY CELLS FOR HANDLING CONCENTRATED SAMPLES IN ANALYTICAL ULTRACENTRIFUGATION

*QUY ONG\*, XUFENG XU and FRANCESCO STELLACCI*

Laboratory of Supramolecular Nanomaterials and Interfaces, Ecole Polytechnique Fédérale de Lausanne (EPFL), Station 12, 1015 Lausanne, Switzerland.

\*Corresponding author

Quy Ong

Laboratory of Supramolecular Nanomaterials and Interfaces, Ecole Polytechnique Fédérale de Lausanne (EPFL), Station 12, 1015 Lausanne, Switzerland.

Email: quy.ong@epfl.ch

Telephone: +4121631002

**ABSTRACT**. In concentrated macromolecular dispersions, far-from-ideal intermolecular interactions determine the dispersion behaviors including phase transition, crystallization, and liquid-liquid phase separation. Here, we present a novel versatile capillary-cell design for analytical ultracentrifugation-sedimentation equilibrium (AUC-SE), ideal for studying samples at high concentrations. Current setups for such studies are difficult and unreliable to handle, leading to a low experimental success rate. The design presented here is easy to use, robust, and reusable for samples in both aqueous and organic solvents while requiring no special tools or chemical modification of AUC cells. The key and unique feature is the fabrication of liquid reservoirs directly on the bottom window of AUC cells, which can be easily realized by laser ablation or mechanical drilling. The channel length and optical path length are therefore tunable. The success rate for assembling this new cell is close to 100%. We demonstrate the practicality of this cell by studying: 1) the equation of state and second virial coefficients of concentrated gold nanoparticle dispersions in water and bovine serum albumin (BSA) as well as lysozyme solution in aqueous buffers, 2) the gelation phase transition of DNA and BSA solutions, and 3) liquid-liquid phase separation of concentrated BSA/polyethylene glycol (PEG) droplets.

KEYWORDS Analytical ultracentrifugation, sedimentation-diffusion equilibrium, equation of states, concentrated dispersion, second virial coefficient, phase transition/separation.

**INTRODUCTION**. In recent years, macromolecules at high concentrations have attracted significant attention in the fields of biochemistry and cell biology. The intermolecular interaction in the interior of cells is very far away from ideality due to high concentrations. For example, more than 20% of the volume for E. coli is occupied by a number of different macromolecules.[1,2] Liquid-liquid phase separation of macromolecules also occurs at high concentrations, forming membraneless organelles inside cells, which play an essential role in the compartmentalization of enzymatic reactions in the cytoplasm.[3] Among techniques capable of measuring the physical states of macromolecules at high concentrations, analytical ultracentrifugation (AUC) is preeminent as a first-principle method to precisely characterize macromolecules in real-time. Especially, AUC-sedimentation equilibrium, AUC-SE, is a popular AUC technique based on the equilibrium between sedimentation and diffusion of the sample, which is well known for its ability to study intermolecular interactions. [4,5,6,7]

An AUC-SE cell usually requires a two-channel centerpiece sandwiched in between two transparent windows made of quartz or sapphire.[7] The centerpiece in some cases can consist of 6- or 8 channels and its thickness is usually in a range from 3.0–12.0 mm. The thickness determines the concentration range usable, the thinner the more concentrated. Indeed, thinner 2-channel centerpieces (down to 1.5 mm) are commercially available for samples that are slightly more concentrated, yet one is still limited to about <10 mg/ml for proteins, and even less for samples that absorb light strongly. Traditionally, AUC-SE is used to measure molecular weights of, for example, proteins, nanoparticles, and supramolecules.[7,8] This application typically requires dilute samples (0.1- 3 mg/ml) for which commercial centerpieces suffice.

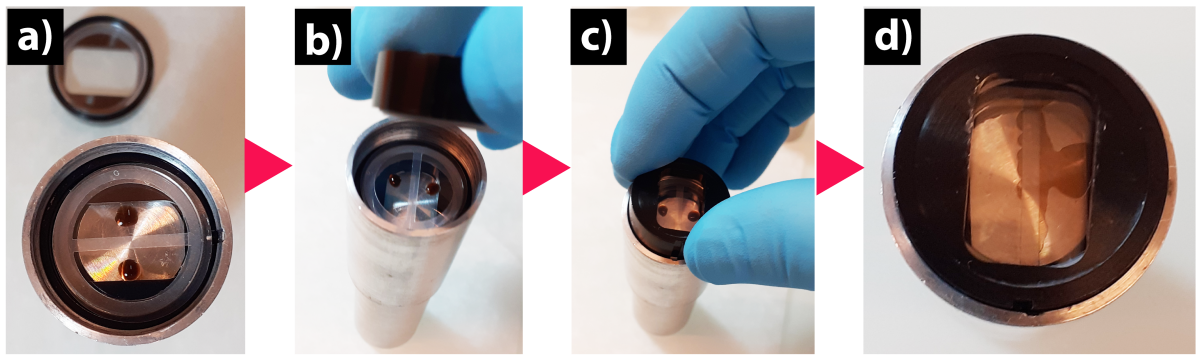

**Figure 1.** Assembly of AUC-SE capillary cell components. Step-by-step assembling (a-c), in which droplets of liquids are deposited onto a bottom window of the cell and topped by a second window before tightening. (d) typical problem of this setup is loss of samples due to capillary force.

Unfortunately, in many recent applications, samples are usually used at high concentrations, for instance, up to 150 mg/ml in a case of therapeutical monoclonal antibodies.[9,10,11] This is also the case for AUC-SE studies that are carried out to understand phase transition and particle-particle interactions, and to study an equation of state via osmotic pressure where concentrated samples are indispensable. In all these cases, regular centerpieces have significant limitations for a number of reasons such as: too low optical transmissivity, high optical aberrations from steep refractive index gradient, and limited range for absorbance. To avoid these problems, it is advantageous to have the centerpieces to be made as thin as possible.[12]

It has been shown previously in a number of publications that replacement of traditional centerpieces with thin double-sector gaskets made of a PTPE or polyester film allowed an optical path length as thin as 40 $\mu$m, **Fig. 1**.[13,14,15,16,17,18,19,20] These double-sector gaskets are commercially available and commonly used, despite not being designed as proper channels. The results obtained from experiments are still correct, thanks to the fact that AUC-SE is a thermodynamic equilibrium method that, in essence, is not influenced by the shape of its cells.[7] For example, Page et al. used

AUC-SE with a double-sector gasket to obtain the equation of state for synthetic clay colloids and to study their phase transition boundaries.[16] Our group also used this technique to study the properties of nanoparticle dispersions such as gold nanoparticles of different sizes and surface coatings.[21] However, we and others have found it problematic to load samples into this kind of cell.[12,21] Current setups are difficult and unreliable to handle, leading to a low experimental success rate for a number of reasons which will be elaborated below.

Usually, to work with this type of AUC-SE cells, low liquid volumes must be used for assembling the AUC cells, and in our experience, ~1.2–1.5 $\mu$l is a suitable volume. The problem lies in the fact that there is not a pre-assembled channel and using a larger amount of liquid will result in smearing and loss of sample due to capillary force during the cell assembly. The lower the surface tension of the solution, the more challenging it is to work with this kind of cell. In addition, all of the cell components are loosely held together during the assembly such that most failure occurs before they are tightly torqued. Therefore, despite the use of low volumes, the success rate of this assembly is still low and requires significant user experience. Additionally, while handling such samples of low volume, the evaporation of the sample leads to unreliable initial concentration.

There are several ways to improve the workflow of this kind of assembly. To use higher liquid volumes, Frisch et al silanized both top and bottom windows to render them hydrophobic, and applied it to investigate molecular ordering transition of single and double stranded DNA.[14,15] Yet, the handling problem remains as one presses the two windows and slide them into the cell housing. For dispersions in organic solvents because of the wettability and evaporation, van Rijseel et al offered a multistep approach that utilized capillaries filled with sample and capped both ends with epoxy and then glued to a metal holder before loading into a low-speed analytical centrifuge cell.[22] Another versatile but sophisticated ultrathin centerpiece was designed by Luigjies et al in which the thin 50–70 µm metal centerpiece was glued directly onto a glass window.[23,24] To make a complete cell, a glass top with two windows was then glued on top, followed by a Teflon cover. Such multiple-cell components are only accessible with microfabrication laboratory equipment while handling such ultrathin centerpieces requires substantial user experience. In summary, current approaches toward the implementation of AUC-SE setup for concentrated dispersions suffer from several drawbacks: they are difficult to handle, only work with small volumes of liquids, require special chemical functionalization, or are not suitable for organic and volatile solvents.

In this paper, we present a simple design that allows an AUC-SE cell setup to be prepared in a quick and robust manner, and at the same time avoid all of the above-mentioned drawbacks. In our approach, two reservoirs, one for the sample and the other for the reference are etched directly into the bottom AUC sapphire window such that desirable volumes of liquid can be deposited without interfering with the cell assembling. The cell is completed with a double-sector gasket and a non-etched sapphire window on top. A light torque is applied to hold the pieces together and at the same time is used to adjust the optical path length. Upon starting the AUC-SE run, the liquids in each channel will be transferred immediately to the bottom of the cells during the acceleration period, making the AUC-SE run exactly on the same principle as before. The assembling is simple and robust, and the windows are reusable, while requiring no special chemical modification to them. This kind of setup is tested to stand even at 60000 rpm, and for a week-long experiment. Notably, our design is simple and suitable to both aqueous and organic dispersions. Moreover, as explained below, both channel length and optical path can be easily tuned.

We illustrate the versatility of our design through a number of applications. Firstly, insights into the interactions of nanoscale objects with themselves and surrounding can be gained from the equation of state and second virial coefficients of concentrated solutions.[25,26] The two thermodynamic parameters are readily available from AUC. Examples are given by gold nanoparticles water and proteins such as bovine serum albumin and lysozyme in aqueous buffers. Secondly, hydrogels serve as exemplary models for simulating densely packed extracellular environments.[27,28] The gelation phase transition of DNA and bovine serum albumin (BSA) solutions is readily observable in concentrated samples via AUC. Our capillary cell can facilitate the exploration of intricate fluid dynamics in these systems. Lastly, comprehending the liquid-liquid phase separation phenomena within concentrated BSA/PEG (polyethylene glycol) droplets is indispensable for the development of stable drug formulations.[29,30,31] Our capillary cell is effectively employed to directly quantify BSA concentrations within the protein-rich phase.

**MATERIALS AND METHODS**. All purchased chemical are used without further purification unless indicated below. Gold nanoparticles were purchased from Tedpella. Lumidot™ CdSe, crystal violet, tryptophan, lysozyme, bovine serum albumin (BSA), thymus calf DNA, polyethyleneglycol (MW = 10000g/mol) were purchased from Sigma Aldrich. BSA and lysozymes were purified to obtain monomers by size-exclusion chromatography (SEC-FPLC) using a HiLoad 26/600 Superdex 200PG (Cytiva®, Switzerland) and HiLoad 26/600 Superdex 75PG (Cytiva®, Switzlerand) column, respectively. Phosphate buffer PBS 1x was purchased from Thermo Fisher Scientific. AUC-SE experiments were performed using an XL-I analytical ultracentrifuge (Beckman Coulter), equipped with absorbance optics. Cell components were purchased Beckman Coulter except the sapphire windows which were acquired from Kyburz Sapphire (Switzerland). Gaskets are available from Beckman Coulter, Nanolytics

(https://www.nanolytics.com/), and SpinAnalytical (https://spinanalytical.com/). In the current study, mainly gaskets from Nanolytics were used. Capillary cells were assembled and run until equilibrium using absorption at radial resolution of 0.003 cm. The final equilibrium scans were recorded at radial resolution of 0.001cm.

**RESULTS AND DISCUSSION**. Shown in **Fig. 2** are the components of our proposed design. It is composed of typical components of an AUC cell, a custom-made bottom spacer (width of 6 mm), and a laser-etched sapphire window. The reservoirs on the bottom window are symmetric to the gliding rail of the AUC cell. One reservoir is for a reference liquid, and the other is for a liquid sample. The dimensions of the reservoirs determine the maximum usable volume, and varying the amount of liquid to be used, its column length changes accordingly. In the presented dimensions (details provided in the Supporting Information), liquid samples can be used up to 5 $\mu$l. The liquid is contained in the reservoirs during the assembly of the cell, so it does not interfere with the assembly routine. Upon acceleration at a low speed, <3000 rpm, the liquid is transferred by centrifugal force to the bottom of the cell where the capillary channels are formed, as shown in **Fig. 2d-e**. The cell height is thus proportional to the amount of liquid put in the reservoir (see Supporting Information, **Fig. S1**). This cell is highly stable and shows no loss of material even at 60000 rpm, in **Fig. 2f**. It is also stable for long running time, even for a week. Demonstrated in **Fig. 2g** is the stability of the cell for 90 hrs, from which one can see the menisci and the absorbance remained the same for the duration of the experiment.

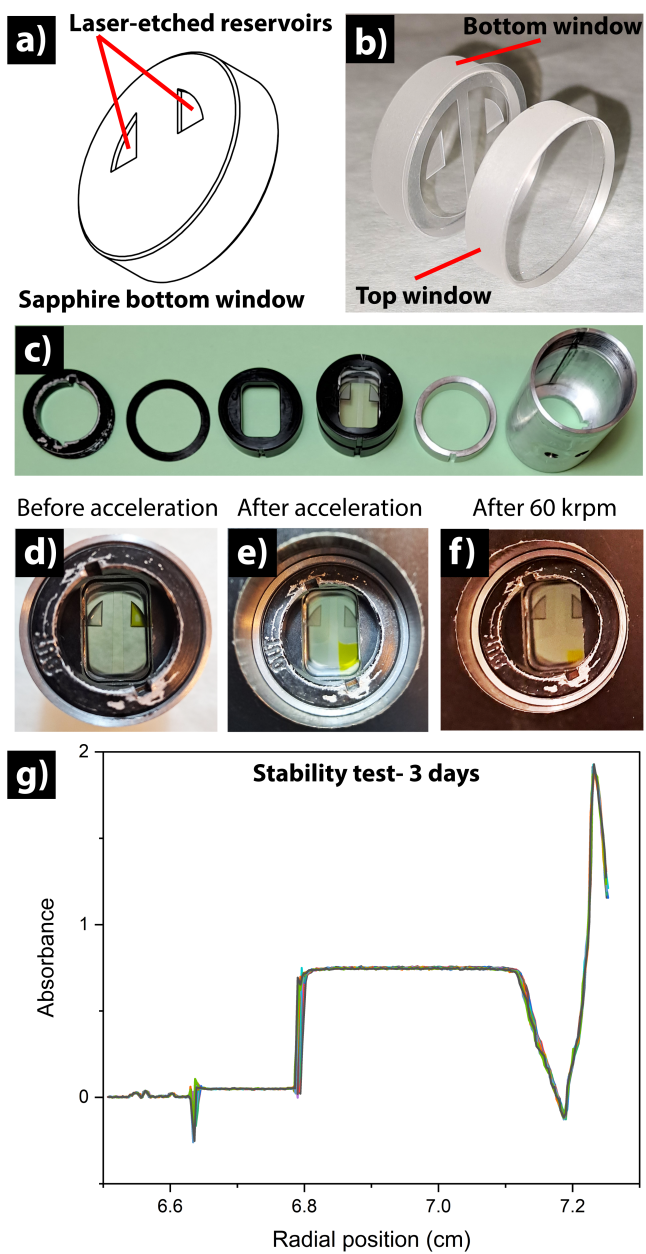

**Figure 2.** New AUC-SE capillary cell design. (a) typical shape of the two laser-etched reservoirs that are fabricated on a bottom sapphire window. (b) the sapphire windows are stacked up with a polyester double-sector gasket to form an ultrathin capillary cell. (c) Components of an AUC-SE cell. (d-f) the visualization of the cell before and after acceleration, and even after 60000 rpm. (g) the sample in this capillary cell can be shown to be stable for a period of 90 hrs.

To show that this design is versatile for samples in both aqueous and organic solvents, we present in **Fig. 3** three different cells that contain tryptophan (2.2 mg/ml) in deionized water, crystal violet (1.5 mg/ml) in ethanol, and commercial quantum dots (5.0 mg/ml) in toluene. All of the samples were successfully assembled for AUC-SE runs, as can be seen in **Fig. 3**, which shows the result of the three setups recorded at 3000 rpm after 3 days of running. The sample crystal violet in ethanol was also ramped up at different rotor speeds up to 60000 rpm. Apart from the well-known radial shift of the sedimentation scans due to the rotor expansion in proportionality to speed, the curves show no loss of materials, confirming the robustness of this design. After the run, visual inspection of the cell showed no crack, or breakage (for example, **Fig. 2f**). After experiments are done, the windows can be washed and reused immediately. The path length of this kind of capillary cell can also be tuned, for example, based on the amount of torque being applied during the tightening of the cell components. Shown in **Fig. S2** are scans of a capillary cell that underwent three different torques, 50, 75, and 100 in. lb, respectively. These corresponded to the path lengths of 97, 87, and 80 $\mu$m that were found based on the optical density.

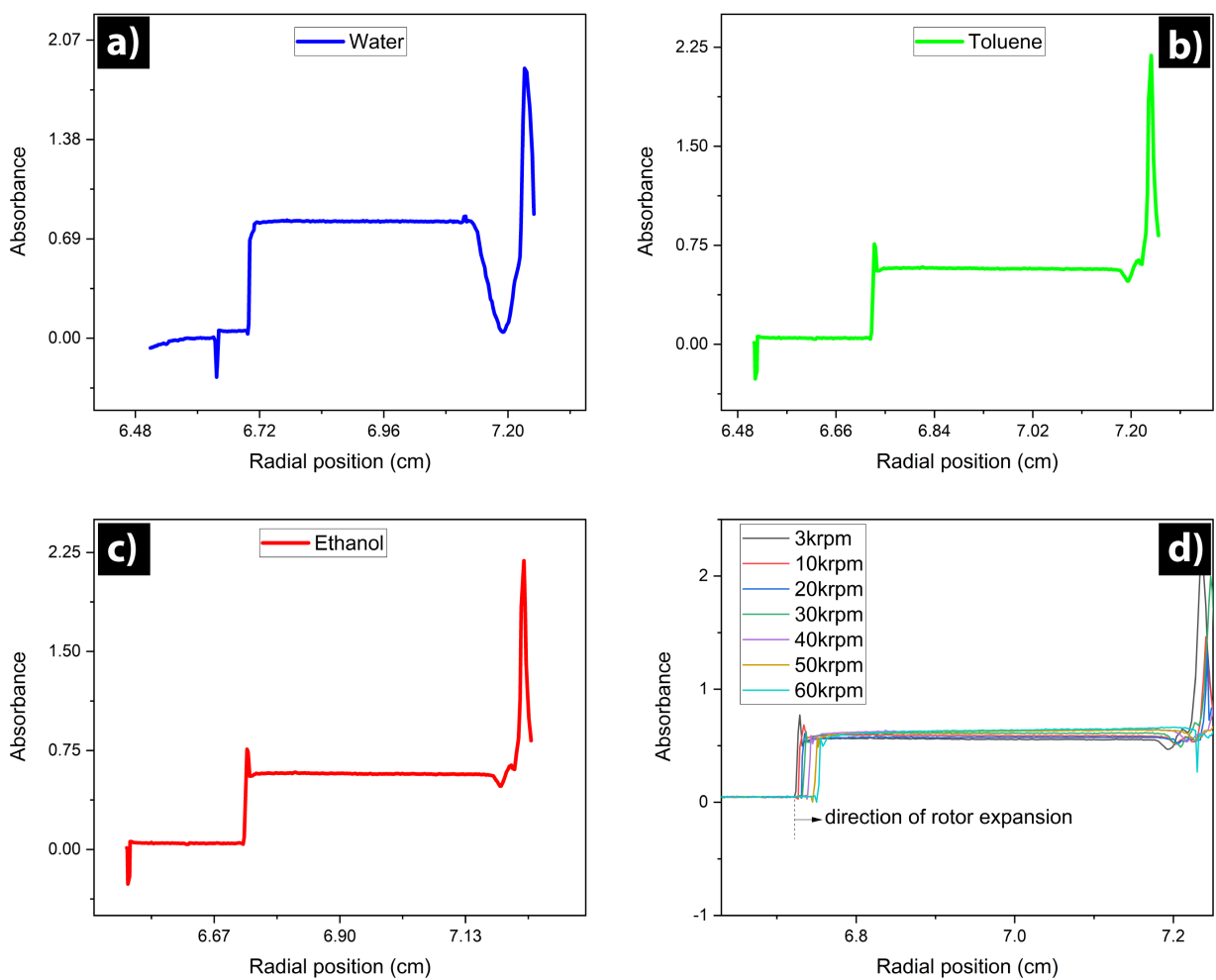

**Figure 3.** Versatility and robustness of new capillary cells for AUC-SE. Examples to demonstrate their use in different solvents (a) tryptophan dissolved in water (2.2 mg/ml), (b) commercial Qdot in toluene (5.0 mg/ml), and (c) crystal violet in ethanol (1.5 mg/ml). (d) The cell withstood a centrifugal force up to the machine's maximum speed (60000 rpm) without showing any cracks or breakage. In figure (d) krpm denotes 1000rpm.

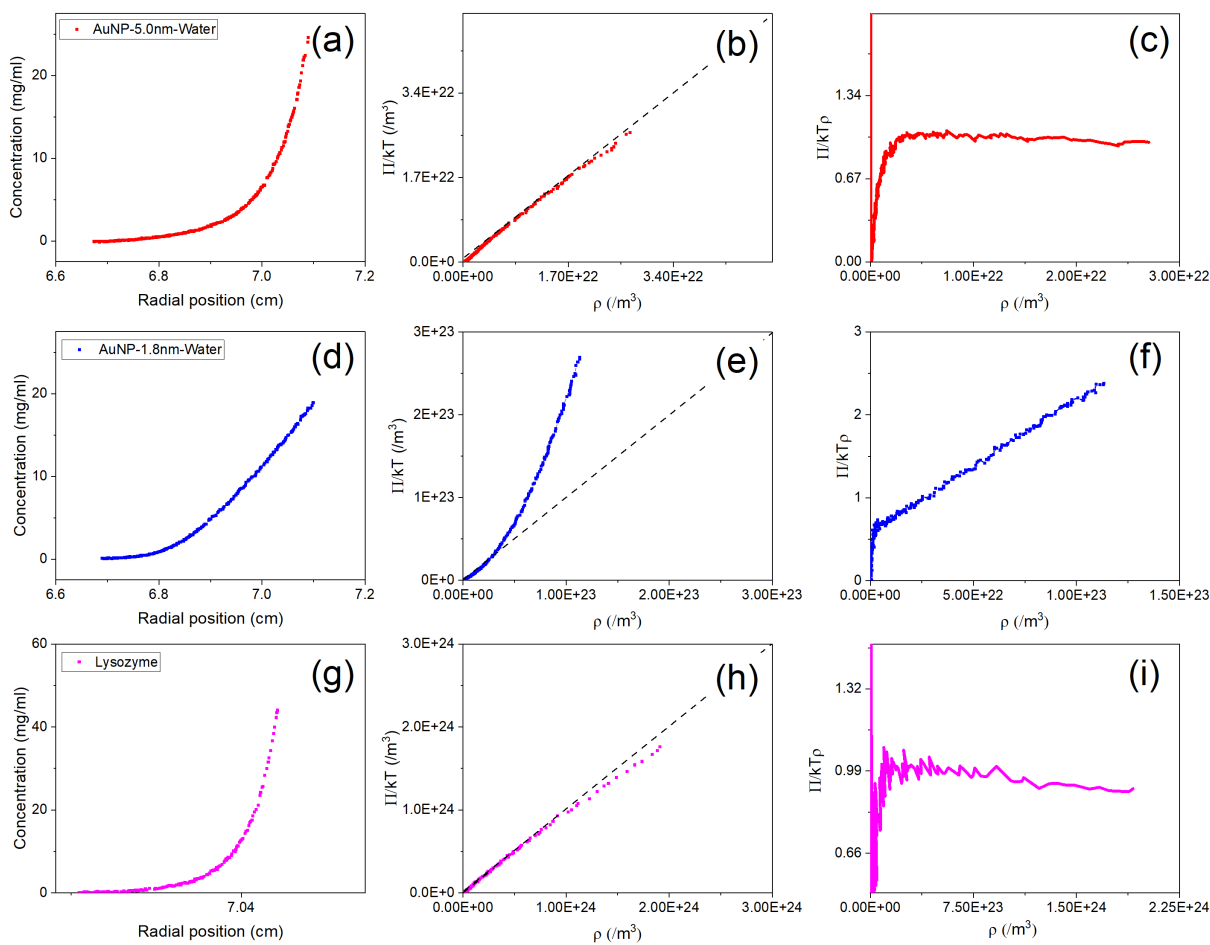

**Figure 4.** Concentration gradient from the sedimentation equilibrium scan (1st column), the calculated EOS curve (2nd column) and the 1st derivative of the EOS curve to calculate the 2nd virial coefficient B22 (3rd column) for (a,b,c) AuNP 5.0 nm in water; (d,e,f) AuNP 1.8 nm in water; (g,h,i) lysozyme in PBS1×.

We applied our capillary cells to study solution non-ideality via the equation of state EOS) and the second virial coefficient ($B_{22}$). We demonstrate it with gold nanoparticles (AuNP) at high concentrations, including AuNP 5.0 nm in water (10 mg/ml), AuNP 1.8nm in water (8 mg/ml) and proteins including BSA and lysozyme. For AuNP, the samples were taken directly from concentrated solutions of the commercial sources. The concentration gradient curves from the SE scans of these samples are shown in **Fig. 4**. The curves were used to calculate the EOS plot for each particle by using the method developed before.[15] Additionally, taking BSA as an example, we show in **Fig. S3** that the EOS of BSA obtained from a typical 3mm centerpiece is similar to that obtained from the current capillary design. By calculating the first derivative of the osmotic pressure in the EOS curves ($\frac{d\pi}{d\rho} = B_{22}\rho$), the second virial coefficients can be thus determined for the measured particle concentration range. For AuNPs 5.0 nm dispersed in water, the EOS curve followed the van't Hoff line (dashed line) in the dilute concentration range which means that in this concentration range, the solution behaves like an ideal solution. However, at the high concentration range, the EOS lies below the van't Hoff line, and we extracted $B_{22}$ from this region and obtained a value of -3.4×10$^{-24}$ m$^3$ suggesting slightly attractive net interaction. For AuNPs 1.8 nm dispersed in water, the EOS curve positively deviates from the van't Hoff line substantially, and the $B_{22}$ = 1.6×10$^{-23}$ m$^3$. For lysozyme dispersed in PBS 1× at 35 mg/ml, we found the value for $B_{22}$ to be -6.5×10$^{-23}$ m$^3$.

The capillary cells are particularly useful for studying phase transformation or equilibrium of phases under centrifugal force. Thymus-calf DNA samples (stock solution at a concentration of 550 ng/$\mu$l) showed gelation by two different centrifugal fields, as shown in **Fig. 5a**. The measurement of the radial scans at the peak absorbance of DNA 260nm provided a limited linear range. Instead, the linear range of the study was extended with a longer wavelength, e.g. 280 nm in **Fig. 5b**, where the non-ideality of the gel is visible. One can see that the gel became compressed significantly toward the bottom of the cell at a higher speed (e.g. 15000 rpm- red scans in **Fig. 5a and 5b**). BSA samples also showed the formation of the gel phase induced by centrifugation, as shown in **Fig. 5c**. The equilibrium between the dilute phase and gel phase is readily visible from the *ln(absorbance)* vs $r^2$ plot as shown in **Fig. 5d**.

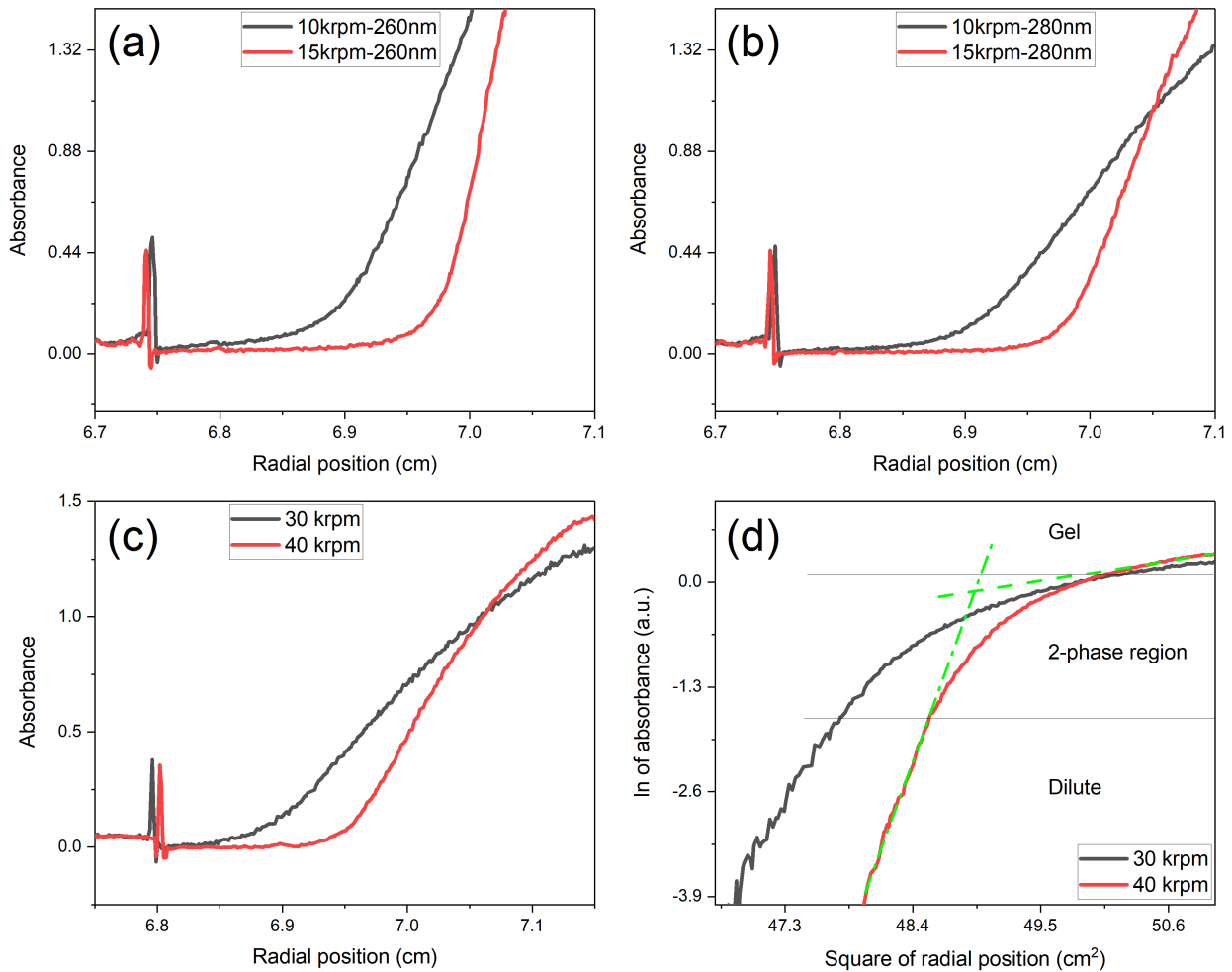

**Figure 5**. Sedimentation equilibrium scans of (a) thymus calf DNA samples by 260 nm at two rotor speeds 10krpm and 15krpm, respectively. Stock solution is 550 ng/$\mu$l measured by Nanodrop (b) of the same run with a wavelength of 280 nm. (c) BSA sample at 30000 rpm and 40000 rpm with a probing wavelength of 280 nm. Stock solution is 114 mg/ml measured by Nanodrop (d) Presentation of the BSA equilibrium scans in a ln(absorbance) vs $r^2$ plot. The dash lines show different linear regions indicative of two distinct phases that are separated by a two-phase-at-equilibrium region. In all legends, krpm denotes 1000 rpm.

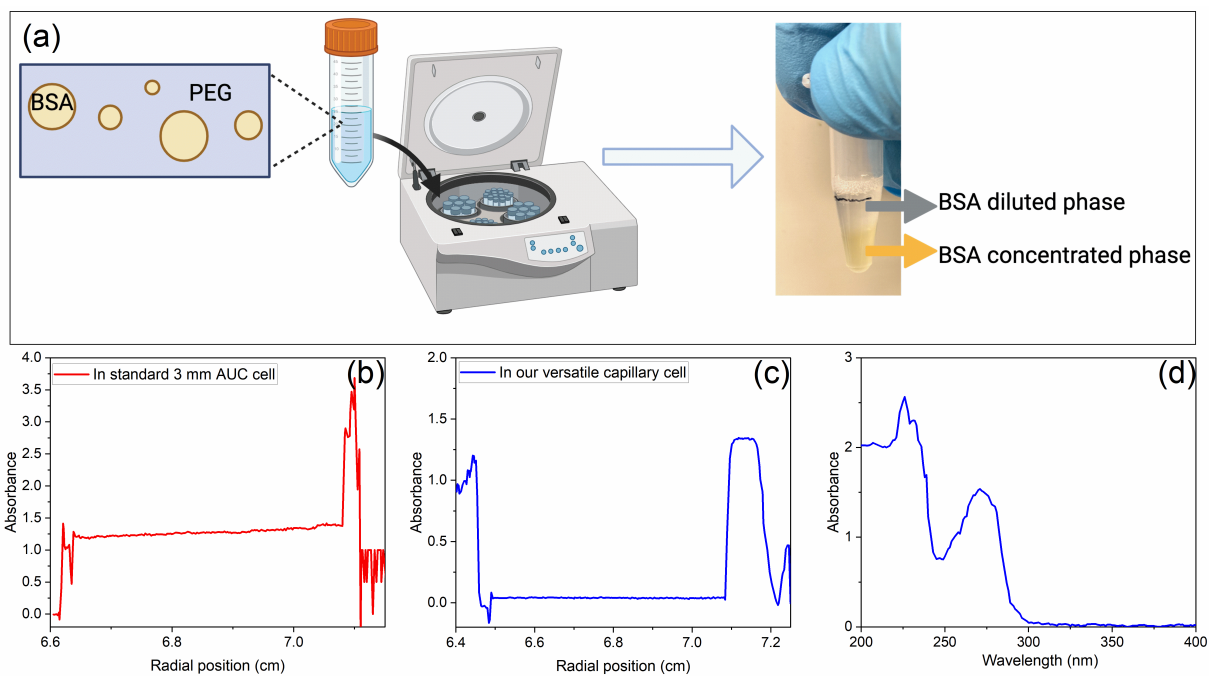

**Figure 6**: Application of new capillary cells for studying liquid-liquid phase separation of BSA droplets in polyethylene glycol (PEG) as a crowding agent. (a) the application of centrifugation to separate the BSA diluted and concentrated phases for measuring the BSA concentration in the two phases; (b) In a standard 3 mm (path length) AUC cell, rotor speed 20000 rpm, the concentration of BSA in diluted phase (from a radial position at 6.64 cm to 7.08 cm) can be measured but the BSA concentrated phase (after 7.08 cm) is too high to measure; (c) In our capillary AUC cell (path length = 87 $\mu$m for torque = 75 in.lb), rotor speed 10000 rpm, the concentration of BSA in concentrated phase (after a radial position of 7.08 cm) is can be measured; (d) the wavelength scan in BSA concentrated phase (from a radial position of 7.08 cm to 7.15 cm) in our capillary AUC cell.

The capillary cell can be also applied in the liquid-liquid phase separation of proteins. In order to construct the phase diagram of proteins, it is crucial to measure the concentration of proteins in both dilute and concentrated phases when the sample is at equilibrium. As shown in **Fig. 6a**, BSA droplets are often separated from the BSA dilute phase by centrifugation. Afterwards, the BSA concentration in the dilute phase can be determined by UV-vis directly while the BSA concentration in the concentrated phase is too high to be measured directly. A conventional practice is to take out a small volume of BSA from the concentrated phase and dilute it with a large volume of buffer. Then the concentration of BSA is low enough to be measured by UV-vis and the BSA concentration in the concentrated phase can be calculated by the dilution ratio. As shown in **Fig. 6b**, the two phases of BSA in a 3 mm cell can be separated by AUC and the concentration of BSA can be measured by UV-vis absorbance optics at 280 nm. The absorbance in the dilute phase (from a radial position of 6.64 cm to 7.08 cm) is 1.25 and the BSA concentration can be thus calculated (= 6 mg/ml). However, the absorbance in the concentrated phase (after the radial position of 7.08 cm) is too high to be measured. In this case, our designed capillary cell can be applied since the path length can be shortened to as thin as 87 μm with torque 75 in. lb. By this means, the absorbance of BSA in the concentrated phase was measured to be 1.34 at 280 nm so that the BSA concentration can be calculated directly (= 233 mg/ml). This setup thus allows the direct measurement of the BSA concentration in the concentrated phase, which will avoid operational errors due to transferring a small volume of the extremely viscous condensate phase.

**CONCLUSIONS**. We presented here a unique design for AUC-SE capillary cells that are robust, reusable, and versatile. They can work with concentrated solutions of various samples in both aqueous and organic medium. We illustrated the application of this design for obtaining equations of state, studying phase separations, and determining the concentration of protein droplets after the phase separation. Overall, we believe the unique capillary cell design will be especially useful for studying concentrated samples by AUC in future.

AUTHOR INFORMATION

**Corresponding Author**

*Quy Ong

Laboratory of Supramolecular Nanomaterials and Interfaces, Ecole Polytechnique Fédérale de Lausanne (EPFL), Station 12, 1015 Lausanne, Switzerland.

Email: quy.ong@epfl.ch

Telephone: +4121631002

**Author Contributions**

QO conceived and designed the AUC cell. QO, XX performed experiments and analysis. FS supervised the project.

**Funding Sources**

SNSF, LIGHT-CAP

**Supporting Information**

Pictures of the windows containing short and long channel designs. Variation of optical path length by mechanical torque force. Comparison of the plots for equation of states of BSA obtained from a standard 3mm cell and a capillary cell. Cell-stand design and dimensions to work efficiently with capillary cells. CAD file for dimensions of capillary cell by laser etching. CAD file for dimensions of the capillary cell by mechanical drilling. CAD file for dimensions of spacer to adjust the capillary centerpiece height.

**Acknowledgement**

SNSF is acknowledged for the support of this project. X.X. and F.S. acknowledge the support of European Union's Horizon 2020 Research and Innovation program under Grant Agreement No. 101017821 (LIGHT-CAP).

ABBREVIATIONS

AUC analytical ultracentrifugation, AUC-SE analytical ultracentrifugation-sedimentation equilibrium; NP nanoparticles; AuNP gold nanoparticles